\documentclass[a4paper, conference]{IEEEtran}
\IEEEoverridecommandlockouts
\usepackage[utf8]{inputenc}
\usepackage{cite}
\usepackage{hhline}
\usepackage{amsmath,amssymb,amsfonts}
\usepackage{algorithmic}
\usepackage{graphicx}
\usepackage{textcomp}
\usepackage{xcolor}
\usepackage[export]{adjustbox}
\usepackage{xurl}
\usepackage{comment}
\usepackage{paralist} 
\usepackage{enumitem} 
\usepackage[bookmarks=false, hidelinks=true]{hyperref}
\usepackage[T1]{fontenc}
\usepackage{tabularx}
\usepackage{multirow}
\usepackage{makecell}
\usepackage{mathtools}

\title{A Risk-Driven Probabilistic Approach to Quantify Resilience in Power Distribution Systems} 

\author{\IEEEauthorblockN{Abodh Poudyal and Anamika Dubey}
\IEEEauthorblockA{Washington State University\\
Pullman, Washington, USA\\
Email: \{abodh.poudyal, anamika.dubey\}@wsu.edu
\vspace{-3em}}

\and
\IEEEauthorblockN{Shiva Poudel}
\IEEEauthorblockA{Pacific Northwest National Laboratory\\
 Richland, Washington, USA\\
Email: shiva.poudel@pnnl.gov 
\vspace{-15pt}}
}

\begin{document}
\maketitle
\thispagestyle{plain}
\pagestyle{plain}

\begin{abstract}
It is of growing concern to ensure resilience in power distribution systems to extreme weather events. However, there are no clear methodologies or metrics available for resilience assessment that allows system planners to assess the impact of appropriate planning measures and new operational procedures for resilience enhancement. In this paper, we propose a resilience metric using parameters that define system attributes and performance. To represent extreme events (tail probability), the conditional value-at-risk of each of the parameters are combined using Choquet Integral to evaluate the overall resilience. The effectiveness of the proposed resilience metric is studied within the simulation-based framework under extreme weather scenarios with the help of a modified IEEE 123-bus system. With the proposed framework, system operators will have additional flexibility to prioritize one investment over the others to enhance the resilience of the grid.
\end{abstract}
\vspace{-15pt}
\section{Introduction}\label{intro}
In recent years, weather-related extreme events have severely affected the performance of electric power systems, especially the aging mid-voltage and low-voltage power distribution grid~\cite{house2013economic}. This calls for proactive threat management of power distribution systems by improving their resilience to high-impact low-probability (HILP) events with the help of new operational procedures and/or hardening of the infrastructure. 
Planning for resilience requires a metric that can not only quantify the impacts of a future event on the grid but also, help in evaluating/comparing different planning alternatives for their contribution to improving resilience~\cite{watson2014conceptual}. 

In literature, multiple articles have sought to define the resilience metrics and have proposed several methods to solve the resilience planning problem. The existing metrics for resilience can be broadly categorized as: a) attribute-based metrics that identify power system attributes such as robustness, resourcefulness, adaptivity, recoverability, and situational awareness~\cite{kandaperumal2020resilience} and b) performance-based metrics that describes the system's ability to maintain supply (i.e., the system's availability~\cite{cai2018availability}) and often measured using the conceptual resilience curve~\cite{panteli2017power}. 
Different resilience indicators that are widely used in literature are based on the optimal repair time of critical components~\cite{wen2020resilience}, energy not served after an extreme event~\cite{espinoza2017seismic}, total critical loads supplied during the aftermath of a disaster~\cite{poudel2018critical}, and  in terms of infrastructure recovery~\cite{umunnakwe2021quantitative}. 
An exhaustive list of examples of resilience indicators for consideration in developing a suitable metric can be found in \cite{gmlc}. 
The resilience of power distribution systems is dependent on several factors such as network configuration, available resources and controls, and several other smart grids features such as distributed energy resources (DERs), smart switches, intentional islanding, and self-healing. Towards this goal, references~\cite{7728107, chanda2016defining} introduce the use of multi-criteria decision-making (MCDM) methods to quantify resilience by taking different topological parameters based on graph theory. 

Despite these existing approaches for defining and quantifying resilience, no formal resilience metric is universally accepted. The existing metrics to quantify power distribution system resilience pose one or more limitations including: (1) they are post-event measures and mostly evaluated for a single event~\cite{gao2016resilience, poudel2018critical, wen2020resilience} ; (2) they do not specifically measure the impacts of HILP events on system performance (kW loss, critical assets without power, total outage duration)~\cite{7728107, chanda2016defining}; (3) they do not provide additional flexibility to system operators to prioritize one investment decision over the other to evaluate the system resilience~\cite{8767980}. Additionally, in resilience assessment events with higher impact and lower probability are considered contrary to reliability analysis~\cite{panteli2015grid}. Thus, we need a probabilistic approach to calculate the resilience metric that captures both the system attributes as well as its response for a given extreme event. 
In prior work, we introduced a framework to evaluate the resilience of power distribution systems using risk-based measures~\cite{poudel2019risk}. Here, we use the framework and evaluate the resilience by combining both the attributes-based and performance-based resilience measures. The combined use will effectively maximize the baseline assessment of resilience and allow system operators to examine different efforts and associated investment activities~\cite{grid2017grid}.
We use a risk-based quantitative measure, conditional value-at risk (CVaR) to derive several resilience-driven parameters that comprise both the system attributes and system performance. These parameters are computed using Monte-Carlo simulation to reflect the impact of risk and uncertainty of extreme events. Finally, a resilience metric is evaluated using the MCDM process that incorporates all of the resilience-driven parameters of the distribution grid.
The specific contribution of this paper are two folds:
\begin{itemize}[noitemsep,topsep=0pt,leftmargin=*]
    \item A novel risk-based resilience metric that considers a comprehensive power system resilience definition. We take multiple resilience-driven parameters -- availability, robustness, brittleness, resistance, and resourcefulness to holistically evaluate the power distribution system resilience based on these parameters. 
    \item A simulation-based approach that allows system operators to evaluate different mitigating actions. The proposed framework provides additional flexibility to prioritize one investment decision over the others to enhance the system's resilience; The operators can come up with economic investment decisions without compromising the resilience of the system.   
\end{itemize}
The rest of the paper is organized as follows: Section~\ref{sec:modeling} details the modeling of extreme events and their impact. Section~\ref{sec:attributes} describes the resilience-driven parameters of a distribution grid. Section~\ref{sec:MCDM} details the MCDM approach using Choquet Integral. Simulation-based Monte-Carlo method is described in Section~\ref{sec:monte-carlo}. Section~\ref{sec:results} demonstrates the effectiveness of the proposed frameworks with results and analysis. Finally, Section~\ref{sec:conclusion} concludes the paper.

\vspace{-2pt}
\section{Event and Impact Modeling}\label{sec:modeling}
\vspace{-3pt}
In this section, we discuss on modeling an extreme event and its impact on the distribution grid. For this work, we only consider wind-related events and their impact which is discussed in the following subsections. 
\vspace{-3pt}
\subsection{Modeling Probabilistic Events}
\vspace{-3pt}
A probabilistic wind event is characterized by the intensity of the wind speed and its probability of occurrence. The intensity here is a function of wind speed, $v$. Although wind-related events have spatiotemporal dynamics~\cite{poudyal2021spatiotemporal}, we assume that for a distribution system, that covers a small region, the wind speed for the entire region is the same. Thus, each of the components in the distribution system experiences a similar wind intensity. The wind speed profile for different intensity levels of the windstorm can be represented by a probability density function (PDF) as discussed in~\cite{poudel2019risk}.
\vspace{-3pt}
\subsection{Line Fragility Models}
\vspace{-3pt}
The impact of the wind-related event can be represented by the fragility model of a distribution line~\cite{7801854}. For simplicity, we only consider the impact of wind-related events on the distribution line and not on the other components of the distribution system. The fragility model of any distribution line gives the outage probability of the line subjected to a particular wind speed and can be represented as:
\vspace{-5pt}
\begin{equation} \small 
\begin{aligned}
&\mathbb{P}_v^l= \begin{dcases}\mathbb{P}_n^l & v < v_{cri} \\
\mathbb{P}_v^l  & v_{cri} \leq v < v_{col} \\
1 & v \geq v_{col} 
\end{dcases} \\
\end{aligned}
\label{eq:outage}
\vspace{-3pt}
\end{equation}
where $\mathbb{P}_n^l$ is the failure rate of line $l$ in normal weather condition, $\mathbb{P}_v^l$ is the failure probability of line $l$ as a function of $v$, $v_{cri}$ is the critical wind speed at which line $l$ experiences failure, and $v_{col}$ is the wind speed threshold beyond which line $l$ is guaranteed to fail.

\section{Resilience of Power Distribution Grid}\label{sec:attributes}
The ultimate goal of a resilient distribution grid is to have a continuous power supply to critical loads (CLs) even during extreme contingencies. In this section we discuss on the resilience curve based on number of CLs and resilience-based parameters of a distribution grid.

\subsection{Resilience Curve}
\begin{figure}[ht]
    \centering
    \includegraphics[width=0.7\linewidth]{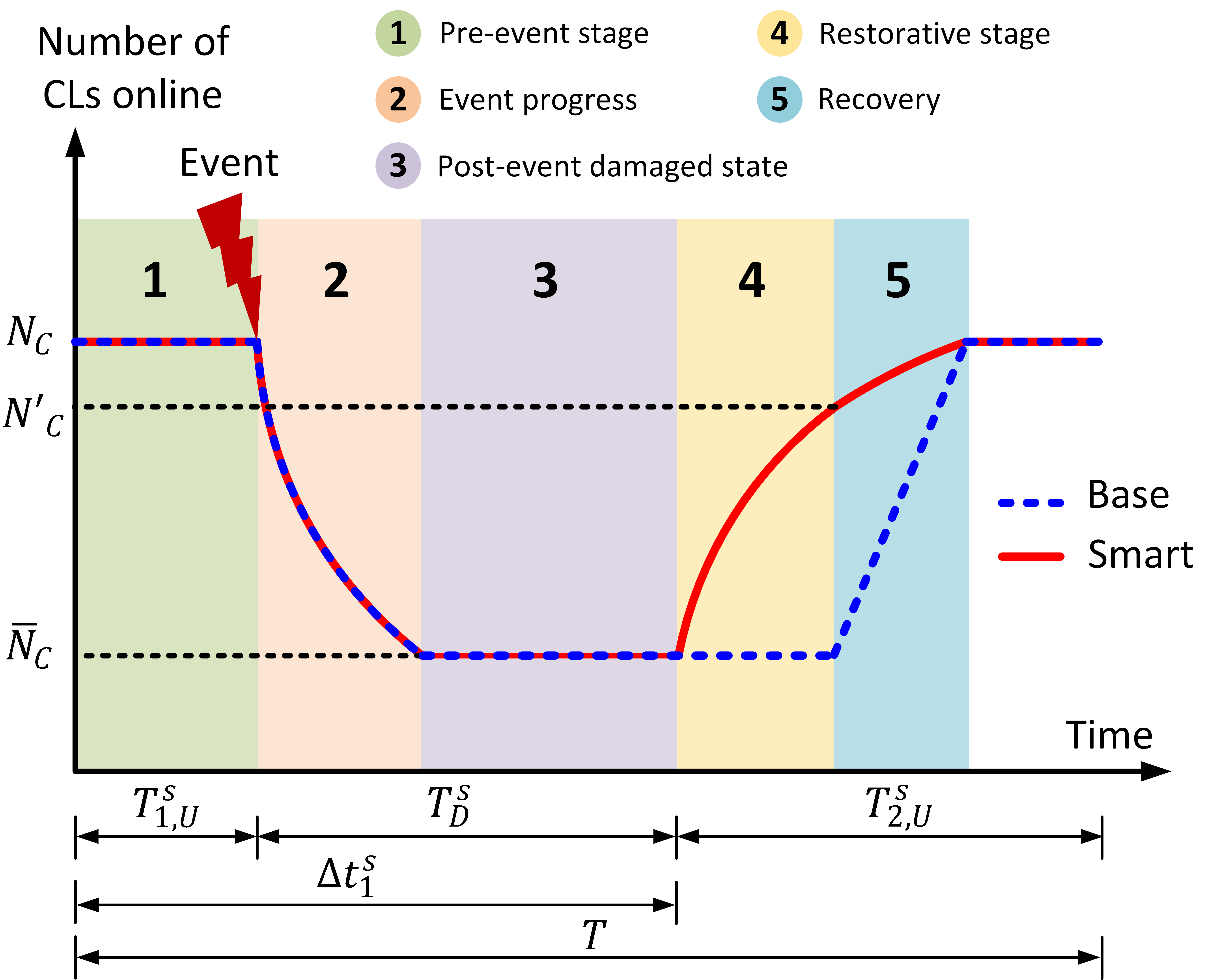}
    \vspace{-3pt}
    \caption{Typical resilience assessment curve based on the number of weighted CL online. The additional time variables on the $x$-axis refer to the smart network.}
    \label{fig:resilience_curve}
    \vspace{-5pt}
\end{figure}

Fig.~\ref{fig:resilience_curve} shows a typical resilience assessment curve in which the $x$-axis represents time whereas the $y$-axis represents the number of weighted CLs online. The plot is represented for two cases namely base network, which does not have any restoration strategy once the event occurs and smart network, in which distributed generators (DGs) and smart damage assessment tools are placed for enhanced situational awareness and restoration. To avoid any confusion, the time variables representing only the smart network are used in Fig.~\ref{fig:resilience_curve} and have the letter $s$ as superscripts to differentiate them from the base network variables. 

Let $N_C$ be the total number of CLs that are online at a particular instance of time. The time these CLs remain online to the time an event occurs is denoted by $T_{1,U}$ and represented by phase 1. It is the time in which all of the CLs remain online. The event occurs at the end of $T_{1,U}$ and sustains for a certain time. The time of event progress depends on the nature and intensity of the event and is denoted by phase 2 of the resilience assessment curve. Some CLs get disconnected due to the severity of the event. $\Bar{N}_C$ be the number of CLs that remain online after an event occurs. Phase 3 denotes the time for damage assessment. Smart networks have smart devices and damage assessment tools that can decrease the damage assessment time significantly. The CLs get disconnected when an event occurs until the point when repair or restoration starts. This is the downtime for CL and is denoted by $T_D$. $\Delta t_1$ is the period from the initial time to the time when repair/restoration begins. For the base case, the repair does not start until the recovery state, phase 5, whereas for the smart network DGs and remote-controlled switches (RCSs) can assist in load restoration, phase 4. At the point of repair/restoration, some of the CLs become online again and remain online for time $T_{2,U}$. Let $N'_C$ be the number of CLs that are online after the load restoration phase. The total up and down time of CL for the entire duration is represented by $T=T_{1, U} + T_D + T_{2,U}$. 
\vspace{-3pt}
\vspace{-3pt}
\subsection{Resilience-driven Parameters}\label{subsec:attributes}
In this work, we only consider phases 1 through 4 for quantifying the resilience metric. We will discuss a few parameters that help us define the resilience of a distribution grid as referred to the critical loads and phases described in Fig.~\ref{fig:resilience_curve}. A detail explanation of these parameters are given in~\cite{2016} while some of them are modified as necessary for this work.

\subsubsection{Availability}
Let $i = 1, 2, ... ,N_C$ be the CLs connected in a system, $T_U^i = T_{1,U} + T_{2,U}$ be the time period when a CL $i$ is connected to system (up time), and $T_D^i$ be the time period when $i$ is disconnected from the system (down time) due to an extreme event. Hence, availability refers to the fraction of time when $i$ is online and is defined as:
\vspace{-3pt}
\begin{equation} \small
    \mathcal{R}_\psi = \frac{\sum_{i=1}^{N_C} T_U^i}{\sum_{i=1}^{N_C} (T_U^i + T_D^i)}
    \label{eq:availability}
\end{equation}

\noindent
Here, $T_U^i$ and $T_D^i$ for each $i$ depends on the type of network. For smart network, some disconnected CLs are restored in phase 4 which increases the overall availability of the system.  

\subsubsection{Robustness}
Let $n_0$ be the number of CLs that are disconnected from the system at a given time. Then the outage incidence, $\theta$ is defined as:
\vspace{-5pt}
\begin{equation} \small
    \theta = \frac{n_0}{N_C}
    \label{eq:max_incidence}
\end{equation}

\noindent
If $N_C - \Bar{N}_C$ be the maximum number of CLs disconnected from the system and $\theta_{max}$ is the maximum outage incidence for a given time, then robustness is defined as:
\vspace{-3pt}
\begin{equation} \small
    \mathcal{R}_\beta = 1 - \theta_{max} = 1 - \frac{N_C - \Bar{N}_C}{N_C} = \frac{\Bar{N}_C}{N_C}
    \label{eq:robustness}
\end{equation}


\subsubsection{Brittleness}
Let $D$ be the percentage of infrastructure damage in the system. For simplicity, we only consider distribution lines as infrastructures in this work. Brittleness is the level of disruption that occurred in the system with respect to damage. For instance, if the damage of a single distribution line affects the entire system then the system is highly brittle. The brittleness of a system with $N_C$ critical loads is defined as:
\vspace{-3pt}
\begin{equation} \small
    \mathcal{R}_\gamma = 100 \times \frac{\theta_{max}}{D}
    \label{eq:brittleness}
\end{equation}

\subsubsection{Resistance}
According to~\cite{2016}, a system has higher resistance if it can withstand extreme events better and can operate the loads for a longer period before getting disconnected. With this notion, a resistant system should have better physical infrastructures, proper damage assessment methods, and situational awareness in case of extreme events. Furthermore, the resistance is also dependent on the nature of the extreme event. The measure of an extreme event to identify system resistance is defined in~\cite{2016}. Based on the measure of the event and time before which the repair and restoration begins, the resistance of a system is given by:
\vspace{-3pt}
\begin{equation} \small
    \mathcal{R}_\xi = \frac{\sigma \sum_{i=1}^{N_C} T_{1,U}^i}{\theta_{max} N_C \Delta t_1}
    \label{eq:resistance}
\end{equation}

\subsubsection{Resourcefulness}
Let $N_{SW}$ be the number of tie-line switches, $N_S$ be the number of generating sources, and $N_P$ be the number of simple paths from each of the sources to CLs after an event has occurred in a network. Then the available resources are useful only if their existence is meaningful in system restoration. Thus, resourcefulness is defined as:
\vspace{-3pt}
\begin{equation} \small
    \mathcal{R}_\delta = \frac{N_P}{(N_{SW} + N_S) \times N_C}
    \label{eq:resourcefulness}
\end{equation}

For the base network, the only available source is the substation so $N_{S} = 1$ for the base case. For the smart network, $N_{S}$ increases as the number of DG increases. However, the resourcefulness decreases if those DGs are not utilized in network restoration after the event has occurred which is ensured by $N_C$. Thus, resourcefulness can be useful for planning the placement and number of DGs to enhance system resilience.

\subsection{Risk-based Resilience Measures}\label{sec:riskcalculation}
A resilient distribution system should not only handle the expected events but also events with lower probability of occurrence that might impose greater impact on the grid. As discussed in~\cite{poudel2019risk}, we use $CVaR$ as risk measure for each of the parameters. $VaR$ is defined as the specific threshold $\zeta$, such that with a specified probability of $\alpha$ $VaR$ does not exceed $\zeta$. On the other hand, $CVaR$ is the expected value of the distribution that exceeds $VaR$. Both of these metrics depend on the value of $\alpha$ and are commonly represented as $VaR_\alpha$ and $CVaR_\alpha$. If $p(I)$ be the probability distribution of a random weather event $I$ then the cumulative probability distribution that the parameter $\mathcal{R}$ will not exceed $\zeta$ when impacted by $I$ is given by:
\vspace{-5pt}
\begin{equation} \small
    \Psi (\zeta) = \int_{\mathcal{R}(I)\leq \zeta }^{} p(I) dI
\end{equation}

\noindent Thus, $VaR_\alpha$ and $CVaR_\alpha$ are then defined by:

\begin{equation} \small
    VaR_\alpha (\zeta) = \inf\{\zeta \in \mathbb{R}:\psi(\zeta)\geq \alpha\}
    \label{eq:var}
\end{equation}
\vspace{-5pt}
\begin{equation} \small
    CVaR_{\alpha} (\zeta) = (1-\alpha)^{-1}\int_{\mathcal{R}(I)\geq VaR_{\alpha} }^{} \mathcal{R}(I)\ p(I)\ dI
    \label{eq:cvar}
\end{equation}

\noindent $CVaR_\alpha$ represents the value of parameter for the extreme $(1-\alpha)\%$ of impacts. It is also to be noted that the distribution of parameters below and above the specified threshold $\zeta$ represent the complete distribution of extreme events with a probability of $\alpha$ and $1-\alpha$ respectively.

\begin{figure*}[t]
    \centering
    \includegraphics[width=0.8\textwidth,trim={0.5cm 0.25cm 0.35cm 0.5cm}, clip] {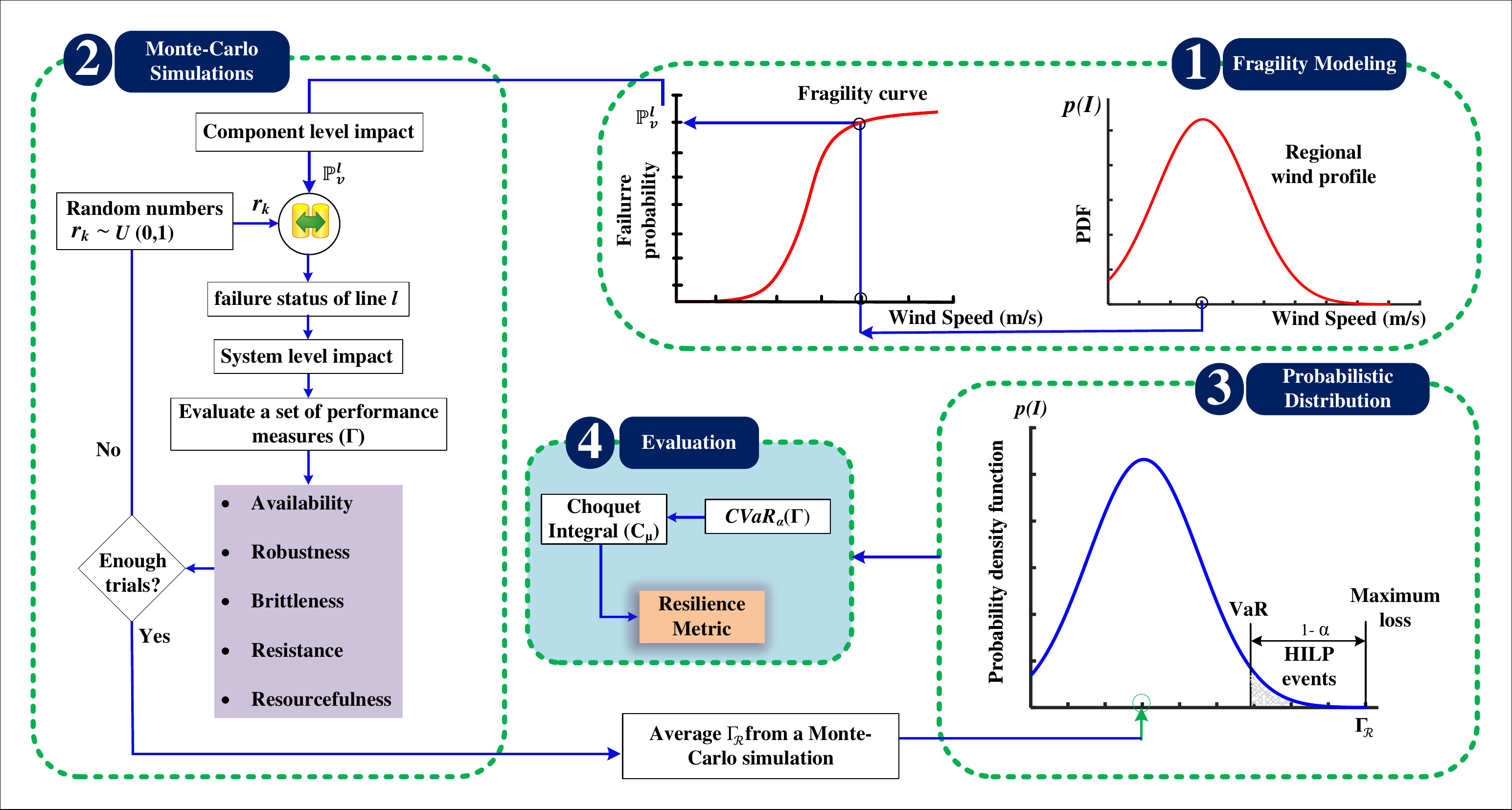}
    \vspace{-0.2 cm}
    \caption{Simulation-based framework for resilience metric computation; Fragility modeling feeds the failure probability to Monte-Carlo simulation. Monte-Carlo simulation calculates the average value of a performance measure for a given event. $CVaR_\alpha$ is then calculated using the pdf of a given extreme event case.}
    \label{fig:ysys}
    \vspace{-0.3 cm}
\end{figure*}

\section{Multi-criteria Decision Making \\using Choquet Integral}\label{sec:MCDM}
All of the parameters defined in Section~\ref{subsec:attributes} are important for enhancing the resilience of a distribution system. Thus, an efficient decision-making strategy is required to identify the important parameters to focus on. Choquet Integral is an effective method for MCDM problem~\cite{choquet} and is well suited for our framework.

\subsection{$\lambda$-Fuzzy Measures}
Let a finite universal set be defined by $\Gamma$, which has $N$ parameters; $\Gamma = \{\mathcal{R}_1,\mathcal{R}_2,...,\mathcal{R}_N\}$. If $\mathbb{P}(\Gamma)$ be the power set of $\Gamma$, then a fuzzy measure on $\Gamma$ is defined by, 
\vspace{-3pt}
\begin{equation} \small
    \mu:\mathbb{P}(\Gamma)\rightarrow[0,1]
\end{equation}
if and only if, a) $\mu(\phi) = 0$, $\mu(\Gamma) = 1$ and b) $A$ $\subset$ $B$ $\subset \Gamma$ $\Rightarrow \mu(A)\leq\mu(B)$. Here, a) ensures that every parameter contributes something and the contribution is maximum when all parameters are included in the set whereas b) shows the monotonic property of fuzzy measures which means that the interaction of parameters should not overshadow the contribution of individual parameters. If $\lambda$ be the interaction degree between two disjoint sets $P$ and $Q$, then for $\lambda > -1$, the Sugeno $\lambda$-fuzzy measure is defined as: 
\vspace{-3pt}
\begin{equation} \small
    \mu(P\cup Q) = \mu(P) + \mu(Q) + \lambda\mu(P)\mu(Q)
\end{equation}
where $\lambda$ is obtained by solving the first condition of fuzzy measures, i.e., $\mu(\Gamma) = 1$. A detail explanation of calculating $\lambda$ is given in~\cite{2006}. 

\subsection{Behavioral Analysis of Fuzzy Measures}
Although $\lambda$ defines some form of interaction among different parameters, the initial fuzzy weights do not provide concrete evidence on the importance of using one criterion over the other. The Shapely index, also known as the importance index, provides insight on interpreting the fuzzy measures~\cite{2006}. For any parameter $\mathcal{R} \in \Gamma$, the Shapely index of $\mathcal{R}$ is defined as:
\vspace{-2pt}
\begin{equation} \small
    \eta_\mathcal{R} := \sum_{\mathcal{S}\subset \Gamma \backslash \mathcal{R}}\frac{(N - |\mathcal{S}| - 1)!|\mathcal{S}|!}{N!} \left[\mu(\mathcal{S}\cup\{\mathcal{R}\}) - \mu(\mathcal{S})\right]
    \label{eq:shapely}
\end{equation}

\noindent
where $|.|$ denotes the cardinality of a set and $\eta_\mathcal{R}$ is the Shapely index of parameter $\mathcal{R}$. The Shapely index is based on the interpretation that the weight of a parameter $\mathcal{R} \in \Gamma$ should not only be defined by its individual fuzzy measure $\mu(\{\mathcal{R}\})$ but by all $\mu(\mathcal{S}\cup\{\mathcal{R}\})$ such that $\mathcal{S}\subset \Gamma \backslash \mathcal{R}$. The term $\mu(\mathcal{S}\cup\{\mathcal{R}\}) - \mu(\mathcal{S})$ is defined as the marginal contribution of parameter $\mathcal{R}$ in $\mathcal{S}$. In this work, $\eta_\mathcal{R}$ is used as the initial weight of each $\mathcal{R}$.

\subsection{Choquet Integral}
If $\mu$ denote the fuzzy measure on $\Gamma$ then the discrete Choquet integral of a function $f:\Gamma\rightarrow\mathbb{R}^+$ with respect to $\mu$ is defined as~\cite{choquet}:
\vspace{-3pt}
\begin{equation} \small
    \mathcal{C}_\mu(f) := \sum_{i=1}^{n}(f(i)-f(i-1))\mu(\{\mathcal{R}_1, \mathcal{R}_2, .... , \mathcal{R}_n\})
    \label{eq:choquet}
\end{equation}

\noindent
where $f(.)$ are arranged in ascending order of its magnitude and is the $CVaR_\alpha$ of the parameters calculated using~(\ref{eq:cvar}), $\mu({\mathcal{R}})=\eta_\mathcal{R}$ is obtained from~(\ref{eq:shapely}), and $f(0) = 0$. Choquet integral gives the overall score of alternative decisions in problem involving multiple parameters for each decision.

\section{Resilience Metric Evaluation Framework}\label{sec:monte-carlo}
In this section, we describe the simulation-based framework to quantify the resilience of power distribution systems. First, each resilience parameter is evaluated using a probabilistic method, and a corresponding risk-based metric is defined. Next, these parameters are combined with Choquet integral that evaluates a single value based on multiple different parameters and their associated importance in the decision-making process. Fig.~\ref{fig:ysys} shows the overall framework to quantify the system resilience using a stochastic simulation-based approach and is described in detail below.

\subsection{Evaluating Resilience-driven Parameters}
The extreme wind event and its impact is characterized using its probability distribution and line fragility model as described in Section~\ref{sec:modeling}. Since the process of identifying an event and its impact is purely stochastic, Monte-Carlo simulations are conducted to evaluate the probabilistic impacts of the event on the power distribution grid. The approach is generic as each event is simulated for several trials. The fragility models provide the failure probability of any distribution lines. With the increase in wind intensity, the failure probability increases accordingly. Monte-Carlo simulations help us identify the number of lines being failed in each trial, and resilience-driven parameters are evaluated using~(\ref{eq:availability}) -- (\ref{eq:resourcefulness}). For smart network, the optimization framework using DGs are modeled and simulated as described in~\cite{poudel2018critical}. At the end of each simulation, the average of evaluated parameters for all trials is then mapped with the respective intensity of the events to form a distribution of each parameter corresponding to its intensity. 

\subsection{Risk-driven Resilience Quantification}
The probability distribution of each of the parameters corresponds to the distribution of the intensity of the event. Thus, $CVaR_\alpha$ of each of the parameters can be calculated using~(\ref{eq:cvar}). It is to be noted that the value of $\alpha$ is consistent for each of the parameters. To combine $CVaR_\alpha$ in the decision-making process, the priorities of each of the parameters are obtained from the system operators and Shapely values of those priorities are evaluated using~(\ref{eq:shapely}). Finally, based on the $CVaR_\alpha$ of each of the parameters and their Shapely values, Choquet Integral gives an overall score using~(\ref{eq:choquet}). To identify the interaction of each of the parameters, $\lambda$ is also considered in the overall calculation process. The overall score obtained from Choquet Integral is the resilience metric for the distribution system. The described process is holistic as it considers all of the resilience-driven parameters (both attribute-based and performance-based) with their priorities in a system along with the associated risk.  

\section{Results and Analysis}\label{sec:results}
\begin{figure}[ht]
    \centering
    \includegraphics[width=0.9\linewidth]{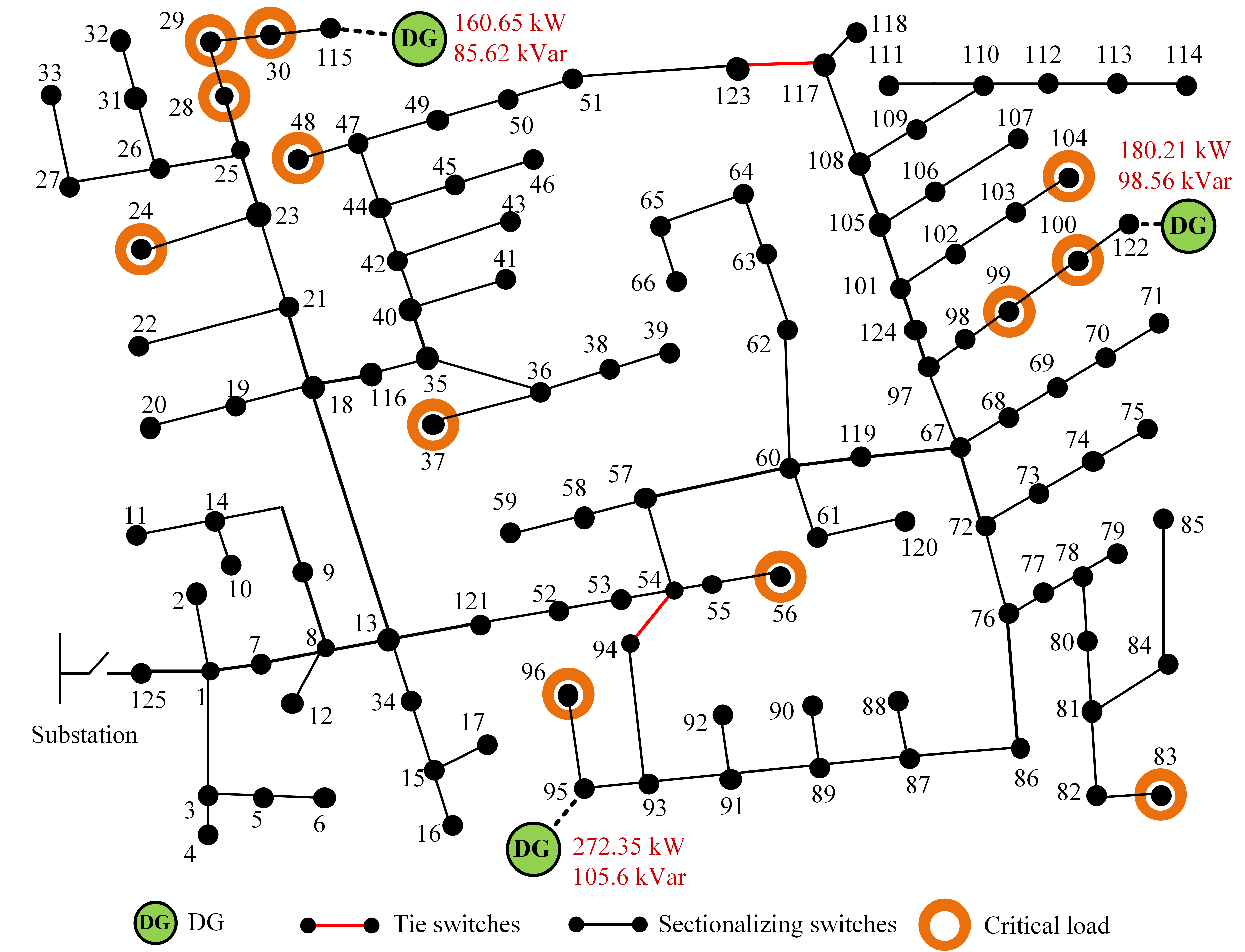}
    \caption{Modified IEEE-123 test case with DGs, tie switches, and CLs.}
    \label{fig:test_case}
\end{figure}

The proposed method of resilience metric quantification using $CVaR_\alpha$ of multiple parameters and Choquet Integral is demonstrated on IEEE 123-bus test system, Fig.~\ref{fig:test_case}. The simulation is carried out for extreme wind-related events. It was experimentally verified that 1000 trials are enough to achieve convergence of MCS for any wind speed scenarios.

\vspace{-5pt}
\subsection{Calculating CVaR of Parameters}
\begin{figure}[t]
    \centering
    \includegraphics[trim=0cm 0.3cm 1.45cm 0.5cm,clip,width=0.8\linewidth]{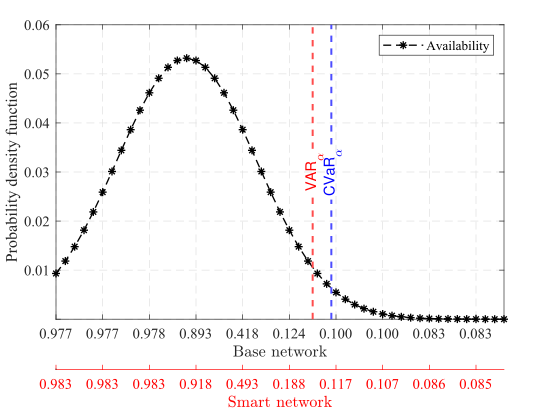}
    \vspace{-5pt}
    \caption{PDF of availability for base and smart network.}
    \label{fig:availability}
    \vspace{-10pt}
\end{figure}

The five parameters defined in Section~\ref{sec:attributes} are calculated based on Fig.~\ref{fig:resilience_curve} and using the simulation method described in Section~\ref{sec:monte-carlo}. Fig.~\ref{fig:availability} shows the PDF of $\mathcal{R}_\psi$ obtained for each wind speed along with $VaR_\alpha$ and $CVaR_\alpha$ values. For all of the cases, the value of $\alpha$ is set to be 0.95. The $VaR_\alpha$ and the $CVaR_\alpha$ are calculated using~(\ref{eq:var}) and~(\ref{eq:cvar}). The risk metrics for other parameters are calculated in a similar fashion and are shown in Table~\ref{tab:CVAR}. Each of the parameters are normalized using min-max normalization technique for generality.

\begin{table}[ht]
    \centering
    \caption{$CVaR_\alpha$ of normalized resilience-based \\ parameters for base and smart network}
    \begin{tabular}{c|c|c|c|c|c}
          & $\mathcal{R}_\psi$ & $\mathcal{R}_\beta$ & $\mathcal{R}_\gamma$ & $\mathcal{R}_\xi$ & $\mathcal{R}_\delta$  \\
    \hhline{======}
         Base & 0.01115 & 0.00012 & 0.01656 & 0.0037 & 0.00005 \\
    \hline
         Smart & 0.01932 & 0.00012 & 0.01656 & 0.0039 & 0.00314 \\ 
    \hline
    \end{tabular}
    \label{tab:CVAR}
\end{table}

\subsection{Quantifying Resilience using Choquet Integral}
To compute the resilience metric based on the multiple parameters and their respective importance, five different cases are developed. For each of the cases, $\mu(.)$ is assigned for each parameter as shown in Table~\ref{tab:weights}. These are the initial fuzzy weights given by the experts or system operators that indicate the priority of one parameter over others.   
\vspace{-5pt}
\begin{table}[ht]
    \centering
    \caption{Initial fuzzy weights of parameters \\ for resilience metric calculation}
    \begin{tabular}{c|c|c|c|c|c}
          & $\mu(\mathcal{R}_\psi)$ & $\mu(\mathcal{R}_\beta)$ & $\mu(\mathcal{R}_\gamma)$ & $\mu(\mathcal{R}_\xi)$ & $\mu(\mathcal{R}_\delta)$  \\
    \hhline{======}
        Case I & 0.9 & 0.25 & 0.15 & 0.6 & 0.85 \\
    \hline
        Case II & 0.6 & 0.5 & 0.45 & 0.5 & 0.6 \\ 
    \hline
        Case III & 0.3 & 0.8 & 0.85 & 0.6 & 0.2 \\ 
    \hline
        Case IV & 0.9 & 0.6 & 0.6 & 0.6 & 0.2 \\ 
    \hline
        Case V & 0.2 & 0.6 & 0.6 & 0.6 & 0.9 \\ 
    \hline
    \end{tabular}
    \label{tab:weights}
\end{table}

Table~\ref{tab:shapely} shows the Shapely value of each of the parameter calculated using~(\ref{eq:shapely}) from their initial fuzzy weights. These values also indicate the marginal contribution of each of the parameters in the respective cases. For instance, in Case I the importance of $\mathcal{R}_\psi$ and $\mathcal{R}_\delta$ are greater than the importance of other parameters. Hence, these two parameters contribute more towards resilience quantification than the others. For different cases, the marginal contribution of each of the parameters differ according to the priority set by the system operator or expert.
\begin{table}[ht]
    \centering
    \caption{Shapely values of each parameters \\ based of their initial weights.}
    \begin{tabular}{c|c|c|c|c|c}
          & $\eta_{\mathcal{R}_\psi}$ & $\eta_{\mathcal{R}_\beta}$ & $\eta_{\mathcal{R}_\gamma}$ & $\eta_{\mathcal{R}_\xi}$ & $\eta_{\mathcal{R}_\delta}$  \\
    \hhline{======}
         Case I & 0.35235 & 0.07617 & 0.04451 & 0.20400 & 0.32294 \\
    \hline
         Case II & 0.23225 & 0.18573 & 0.16404 & 0.18573 & 0.23225 \\ 
    \hline
         Case III & 0.09441 & 0.30385 & 0.33202 & 0.20849 & 0.06121 \\ 
    \hline
         Case IV & 0.34422 & 0.19903 & 0.19903 & 0.19903 & 0.05869 \\ 
    \hline
         Case V & 0.05869 & 0.19903 & 0.19903 & 0.19903 & 0.34422\\
    \hline
    \end{tabular}
    \label{tab:shapely}
\end{table}

The Choquet Integral values for the base and the smart network and each of the cases are shown in Table~\ref{tab:CI_resilience}. It can be seen that the resilience of the smart network is always greater than that of the base network regardless of individual test cases due to the presence of DGs. However, the resilience for individual networks varies with the Shapely values of each of the parameters. For instance, if we look at the smart network, Case IV is more resilient than any of the other cases as higher priority is given to load availability and infrastructural investments (i.e., $\mathcal{R}_\beta$, $\mathcal{R}_\gamma$, and $\mathcal{R}_\xi$). However, this is not true for the base network as loads are not picked up during the restoration phase in the base network making its availability lower than the smart network. It is also interesting to notice that, the resilience for Case I and Case IV does not have a huge difference although for Case I, the priority towards infrastructural investment is less. Hence, the operators can have the flexibility to focus more on less expensive decisions and still enhance the system's resilience.       
\vspace{-5pt}
\begin{table}[ht]
    \centering
    \caption{Choquet Integral values based on \\ Shapely values of each parameters}
    \begin{tabular}{c|c|c|c|c|c}
        Network & Case I    &   Case II   & Case III & Case IV & Case V  \\
    \hhline{======}
        Base  & 5.45 & 6.03 & 7.36 & \textbf{7.89} & 4.72 \\
    \hline
        Smart & 9.36 & 8.68 & 8.36 & \textbf{10.93} & 6.31 \\ 
    \hline
    \end{tabular}
    \label{tab:CI_resilience}
\end{table}

\section{Conclusion}\label{sec:conclusion}
In this paper, a risk-based resilience metric is proposed which incorporates multiple parameters of the distribution grid that can alter the resilience of the grid. A stochastic simulation-based approach is presented to quantify the resilience of the distribution grid. Since resilient systems should be able to withstand extreme events that have a minimum probability of occurrence, the $CVaR_\alpha$ of the grid parameters for extreme event cases are used to calculate the resilience metric. The simulation results showed that prioritizing one parameter over the others can either enhance or degrade the system's resilience depending upon how the investment decisions are prioritized. Additionally, it was concluded that the framework provides added flexibility to choose economically feasible investments without compromising the resilience of the system. A future avenue of research is to include various planning decisions such as line hardening, DG placements and sizing, and so forth within the optimization framework to quantify the system's resilience.

\bibliographystyle{IEEEtran}
\bibliography{references} 
\end{document}